# Wigner-molecularization-enabled dynamic nuclear field programming

Wonjin Jang[1], Jehyun Kim[1], Jaemin Park[1], Gyeonghun Kim[1], Min-Kyun Cho[1], Hyeongyu Jang[1], Sangwoo Sim[1], Byoungwoo Kang[1], Hwanchul Jung[2], Vladimir Umansky[3], and Dohun Kim[1]*

[1]*Department of Physics and Astronomy, and Institute of Applied Physics, Seoul National University, Seoul 08826, Korea*

[2] *Department of Physics, Pusan National University, Busan 46241, Korea*

[3]*Braun Center for Submicron Research, Department of Condensed Matter Physics, Weizmann Institute of Science, Rehovot 76100, Israel*

*Corresponding author: dohunkim@snu.ac.kr

**Abstract**

Multielectron semiconductor quantum dots (QDs) provide a novel platform to study the role of Coulomb correlations in finite quantum systems and their impact on many-body energy spectra. An example is the formation of interaction-driven, spatially localized electron states of Wigner molecules (WMs). Although Wigner molecularization has been confirmed by real-space imaging and coherent spectroscopy, the open system dynamics of the strongly-correlated states with the environment are not yet well understood. Here, we demonstrate efficient control of spin transfer between an artificial three-electron WM and the nuclear environment in a GaAs double QD. A Landau–Zener sweep-based polarization sequence and low-lying anti-crossings of spin multiplet states enabled by Wigner molecularization are utilized. An efficient polarization rate of 2.58 $h$·kHz·$(g^*\mu_B)^{-1}$ per electron spin flip and, consequently, programmable nuclear polarization by controlled single-electron tunneling are achieved. Combined with coherent control of spin states, we achieve control of magnitude, polarity, and site dependence of the nuclear field. It is demonstrated that the same level of control cannot be achieved in the non-interacting regime. Thus, we confirm the multiplet spin structure of a WM, paving the way for active control of newly emerging correlated electron states for application in mesoscopic environment engineering.

Semiconductor quantum dot (QD) systems facilitate investigations of the interaction between electron spins and nuclear environments, which is known as the central-spin problem[1,2]. Although the fluctuation of nuclear fields, which is quantified by the effective Overhauser field $B_{nuc}$ [3,4], often acts as a magnetic-noise source for spin qubits[3], the hyperfine electron–nuclear spin interaction allows to achieve dynamic nuclear polarization (DNP)[5–8]. DNP is used for enhancing the signal-to-noise ratio in nuclear magnetic resonance[6] and prolonging coherence times in QD-based spin qubits[9,10]. Gate-defined semiconductor QDs have been used to achieve the fast probing of nuclear environments[8,11,12], bidirectional DNP[11], and active feedback control of nuclear fields[10].

Although DNP achieved by the pulsed-gate technique is more relevant for quantum information applications compared to spin-flip mediated transport with an applied bias[13,14], spin qubit control combined with DNP has been limited to two-electron singlet–triplet (ST$_0$) spin qubits[9–12,15]. Despite the versatility of gate-defined QD systems[16–19], the large singlet-triplet energy splitting $E_{ST}$ (~$10^2$ $h$·GHz; $h$ is Planck's constant) in particular in GaAs limits the access to higher spin states[20] in multielectron QDs at moderate external magnetic fields $B_0 < 1$ T or within a typical frequency bandwidth of experimental setups.

Coulomb-correlation-driven Wigner molecules (WMs) in confined systems[21–24] may provide new directions for expanding nuclear control to multielectron systems. Recent studies on QDs in various systems have shown clear evidence of WM formation[22,23,25–28]. It has been demonstrated that the $E_{ST}$ can reach down to $\sim 10^0$ $h$·GHz upon the WM formation[25,27] because of strong electron-electron interactions confirmed by full-configuration interaction (FCI)-based theories[23,28,29]. However, most studies have focused on the spectroscopic confirmation of WM formation, and studies on the open system dynamics using correlated states have not been reported to date.

Here, we demonstrate the formation of a WM in semiconductor QDs, which helps achieving efficient spin environment control. We use a gate-defined QD in GaAs and exploit the quenched energy spectrum of the WM ($E_{ST} \sim 0.9$ $h$·GHz) to enable mixing between different $S_z$ subspaces within $B_0 < 0.5$ T, where $S_z$ denotes the spin projection to the quantization axis. Furthermore, we demonstrate DNP by pulsed-gate control of the electron spin states. Leakage spectroscopy and Landau–Zener–Stuckelberg (LZS) oscillations confirm a sizable bidirectional change in $B_{nuc} \sim 80$ mT and the spatial Overhauser field gradient $\Delta B_{nuc} \sim 35$ mT due to the long nuclear spin diffusion time $\tau_N \sim 62$ s. Further, we demonstrate on-demand control of $B_{nuc}$ combined with coherent LZS oscillations, providing a new route for realizing programmable DNP using correlated electron states.

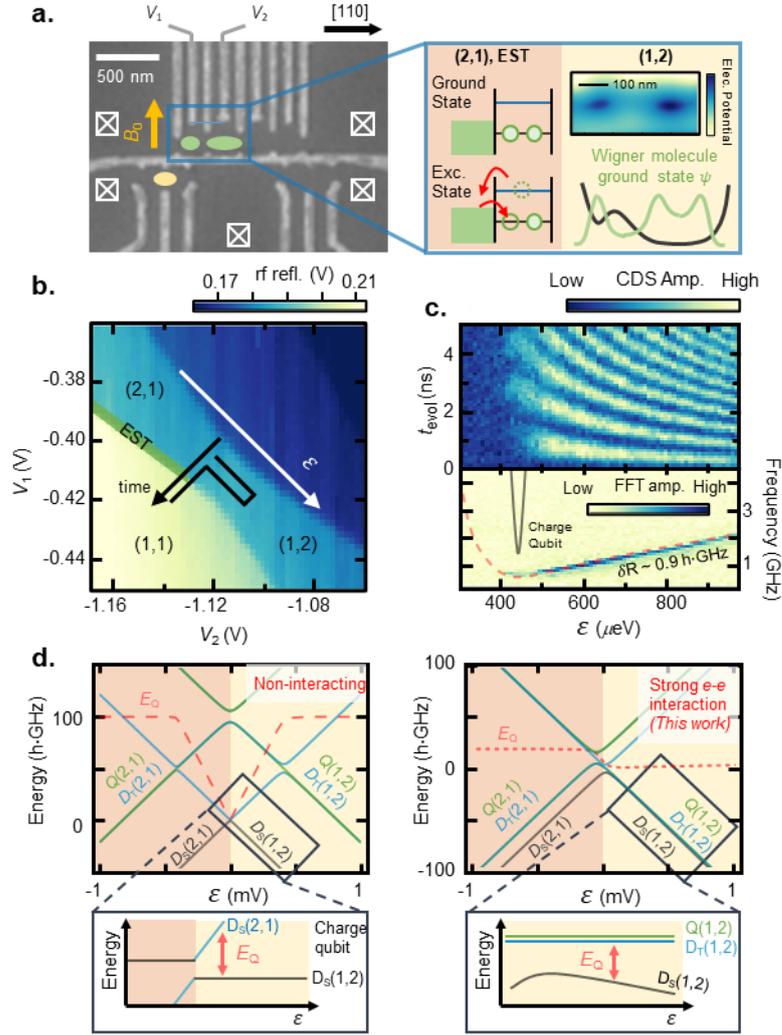

**Fig. 1.**

Figure 1a shows a gate-defined QD device fabricated on a GaAs/AlGaAs heterostructure, where a 2D electron gas (2DEG) is formed ~70 nm below the surface (see Methods). We focus on the left double QD (DQD) containing three electrons. We designed the $V_2$ gate to form an anisotropic potential, which is predicted to promote WM formation[22]. An electrostatic simulation of the electric potential at the QD site near $V_2$ shows an oval-shaped confinement potential with anisotropy exceeding 3 (Fig. 1a, right panel). This potential can be tuned by the gate voltage, allowing the controlled electron correlation and localization of the ground state wavefunction within the DQD[22,24,26,27]. The yellow dot in Fig. 1a. denotes a radio-frequency single-electron transistor (rf-SET) charge sensor utilized for quantum state readout[30–32]. The device was operated in a dilution refrigerator with a base temperature of ~40 mK, an electron temperature $T_e$ ~150 mK (Supplementary Note 1), and a variable $B_0$ applied to the direction shown in Fig. 1a.

First, we show the spectroscopic evidence of the WM at $B_0 = 0$ T by probing $E_{ST}$ in the right QD $\delta R$. Fig. 1b shows a charge stability diagram. The green-shaded region near the (2,1)–(1,1) charge transition is exploited for energy-selective tunneling (EST) readout and state initialization[27,33,34]. We tune the electron tunneling-in (-out) time $\tau_{in}$ ($\tau_{out}$) of the left dot to 14 (7) $\mu$s. Starting from the initialized ground doublet state $D_S$ in the (2,1) charge configuration, we apply non-adiabatic pulses (Fig. 1b) simultaneously to $V_1$ and $V_2$ with a rise time of ~500 ps and a repetition period of 51 $\mu$s $\gg \tau_{in}$ to induce coherent LZS oscillation[35,36]. The oscillation reveals the relative phase evolution between the excited and ground doublet states ($D_T$ and $D_S$), the frequency of which is governed by $\delta R$.

Fig. 1c shows the resultant LZS oscillations as a function of evolution time $t_{evol}$ and detuning $\varepsilon$. The $E_{ST}$ in GaAs DQDs in the non-interacting regime is typically on the order of $10^2$ $h \cdot$GHz[20] (Fig. 1d). In a charge qubit regime, a steep rise in the LZS oscillation frequency $f_{LZS}$ as a function of $\varepsilon$ (Fig. 1c, black curve) and short coherence time $T_2^*$ ~ 10 ps due to strong susceptibility to charge noise is expected[37]. However, we find a significantly smaller $f_{LZS}$ in the (1,2) charge configuration and $T_2^*$ ~ 10 ns because of the reduced dispersion of $f_{LZS}$ versus $\varepsilon$. This is a reminiscent of a QD hybrid qubit[27,36,38], but the excited energy is suppressed owing to the electron–electron interaction. WM formation in our previous GaAs device has been recently confirmed by FCI calculation[27–29]. Although such calculation is needed to rigorously determine parameters, we roughly estimate $\delta R$ ~ 0.9 $h \cdot$GHz, by fitting the fast Fourier transformed (FFT) spectrum to the calculation result (Fig. 1c, red-dashed curve) derived from a toy-model Hamiltonian[33,35,36] (see Methods).

The full energy spectrum calculation of the three-electron states using the parameters obtained experimentally across the (2,1)–(1,2) configuration is illustrated in Fig. 1d (right panel). The suppressed $E_{ST}$ of the left dot $\delta L \sim 19$ $h$·GHz is obtained by measuring the width of the EST region in the charge stability diagram with the lever arm of the gate $V_1 \sim 0.03$. Because of the small value of $\delta L/(k_B T_e) \sim 6$, where $k_B$ is Boltzmann's constant, thermal tunneling precludes high-fidelity single-shot readout. We obtain data by the time-averaged signal using the correlated-double sampling (CDS) method, which effectively yields the signal proportional to the excited state probability[33] (see Supplementary Note 2).

We confirm the WM spin structure via the strongly suppressed energy spectrum in the right QD with varying $B_0$. We focus on five low-lying energy levels among eight possible multiplet states. See Methods for notations used for labeling spin multiplets. Hereinafter, $n$ ($m$) denotes the number of electrons in the left (right) dot by ($n$, $m$; $S_z$). As Fig. 2a (left panel), $D_S(1,2;-1/2)$ ($D_S(1,2;1/2)$) becomes degenerate with $D_T(1,2;1/2)$ or $Q(1,2;1/2)$ ($Q(1,2;3/2)$) at a certain $\varepsilon$ depending on the $B_0$ magnitude. The degeneracies are lifted by the transverse Overhauser field $B_{nuc}^{\perp}$ [8,11]. To detect such anti-crossings, we first initialize the state to either $D_S(2,1;-1/2)$ or $D_S(2,1;1/2)$ at the EST position. By pulsing the initialized $D_S(2,1;-1/2)$ ($D_S(2,1;1/2)$) towards (1,2) and holding for ~100 ns $\gg T_2^*$, mixing with (or leakage to) states $Q(1,2;1/2)$ or $D_T(1,2;1/2)$ ($Q(1,2;3/2)$) can occur if the pulse amplitude $A_P$ coincides with the anti-crossing position (Fig. 2a, right panel). Upon pulsing back to the (2,1) charge configuration, the resultant excited states $Q$ or the $D_T$ probability can be detected via EST[27,33,34]. Fig. 2b shows the leakage spectrum versus $A_P$ and $B_0$, mapping out the anti-crossing positions similar to "spin-funnel" measurements in two-electron $ST_0$ qubits reproducing the energy splittings between the ground and excited levels[8,16,39,40]. The black (red) dashed curves show the calculated splittings (Fig. 1d) between the $D_S$ and $D_T$ ($Q$) states at $B_0 = 0$ T, with the Lande $g$-factor $g^* \sim -0.4$ [41,42].

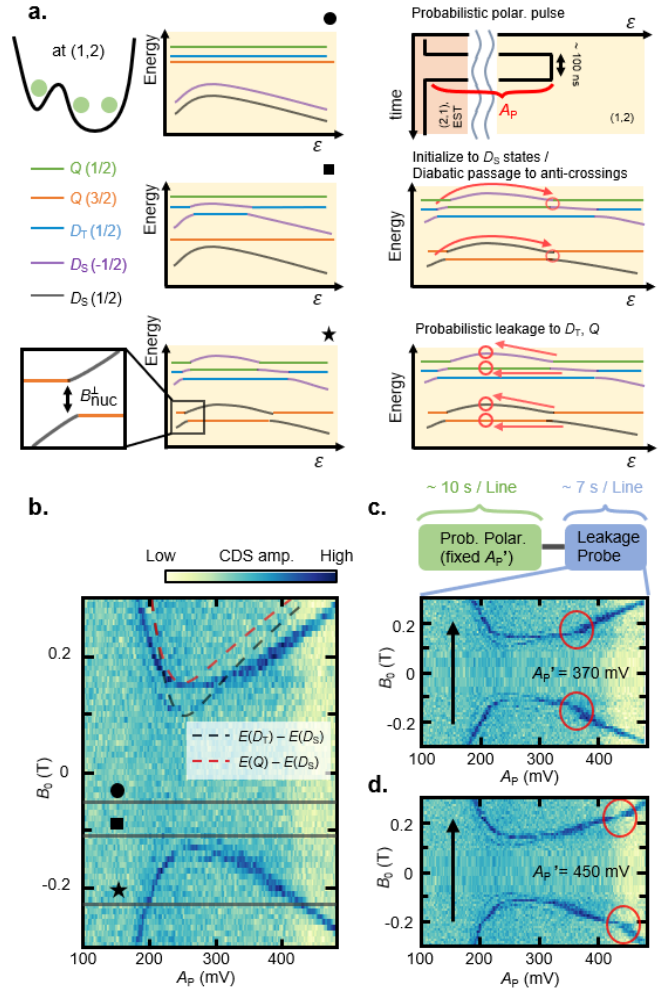

**Fig. 2.**

Although the calculated curve qualitatively agrees with the experimental curve, the observed spectrum curvature as a function of $A_P$ and $B_0$ is smaller because of the DNP induced by the pulse sequence used for leakage spectroscopy. To confirm this, before each line scan of $A_P$ in Fig. 2c (2d), a similar step pulse with a fixed amplitude $A_P' \sim 370$ mV (450 mV) is applied for 10 s. Consequently, we observe distortions (red circles) in the spectrum occurring at $A_P'$.

This is because, when $A_P'$ matches with the anti-crossing position, the pulse probabilistically flips the electron spin with a change in the angular momentum $\Delta m_S = +1$ by the leakage process described above and accompanies flop $\Delta m_N = -1$ of the nuclear spin[8,11]. Unlike the electrons in GaAs, nuclei have positive $g$-factors [8,20]; therefore, the pulse polarizes $B_{nuc}$ toward the $B_0$ direction. This additionally drags the leakage spectrum opposite to the $B_0$ direction under a specific condition $A_P = A_P'$. These results indicate that leakages induced by hyperfine interaction between the WM and nuclear environment lead to an observable change in $B_{nuc}$. Despite the long measurement time per line scan (~7 s) owing to the communication latency between the measurement computer and the instruments, the polarization effect is still visible. Thus, $\tau_N > 10$ s, as discussed below. Moreover, as the anti-crossing position is a sensitive function of $B_{tot} = B_0 + B_{nuc}$ over 100 ~ 300 mT, it can be used to measure $B_{nuc}$.

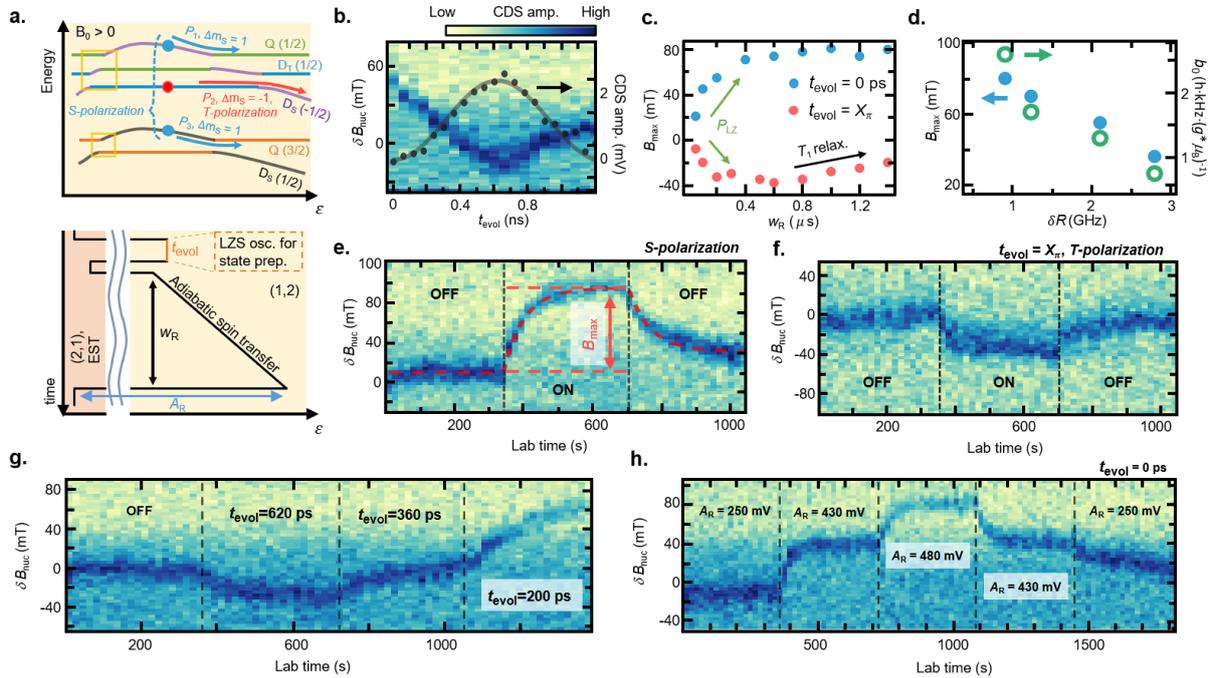

**Fig. 3.**

We now show bidirectional DNP combined with coherent control of doublet states at $B_0 = 230$ mT. Fig. 3a (top panel) shows the three primary paths through the anti-crossings, which can flip the electron spins deterministically by adiabatic passage[2,8,11]. Paths $P_1$ and $P_3$ describe the S-polarization that flips the electron spin with $\Delta m_S = +1$. This is enabled by initializing the state to $D_S(1,2;-1/2)$ ($D_S(1,2;1/2)$) at the EST position and then by non-adiabatically pulsing beyond the first anti-crossings near the (2,1) charge configuration (Fig. 3a, yellow boxes), followed by adiabatically driving the state through the anti-crossing to $Q(1,2;1/2)$ ($Q(1,2;3/2)$), which accompanies $\Delta m_N = -1$ (Fig. 3a, blue arrows). The $Q(1,2;1/2)$ ($Q(1,2;3/2)$) state is diabatically driven back to the EST position, and one electron quickly tunnels out to the reservoir. Reloading an electron from the reservoir reinitializes one of the $D_S$ states completing the polarization cycle. Both the $D_S(1,2;-1/2)$ and $D_S(1,2;1/2)$ initial states contribute to the S-polarization. Path $P_2$ denotes the T-polarization ($\Delta m_S = -1$, $\Delta m_N = +1$), which is possible by driving $D_T(1,2;1/2)$ adiabatically to $D_S(1,2;-1/2)$ (Fig. 3a, red arrow). To prepare $D_T(1,2;1/2)$, we apply a π-pulse to $D_S(2,1;1/2)$ before the adiabatic passage (Fig. 3a, bottom panel). The T-polarization is possible only when the state is initialized to $D_S(2,1;1/2)$ at the EST position.

Combining the S- and T-polarizations, we measure the change in $B_{nuc}$ ($\delta B_{nuc}$), where the repeated polarization pulse sequence (Fig. 3a, bottom panel) with variable $t_{evol}$ and a repetition rate of ~ 20 kHz is applied for 10 s before each line scan. For Fig. 3b, a waiting time ~10 min was added after each sweep to allow the polarized nuclei to diffuse and minimize the polarization effect in the next sweep. As shown in Fig. 3b, $\delta B_{nuc}$ oscillates with $t_{evol}$, which is anti-correlated with the LZS oscillation that represents the population of $D_T(1,2;1/2)$. This confirms that the net polarization rates can be controlled by adjusting $t_{evol}$. Accordingly, we calibrate $t_{evol} = 0$ (0.62 ns) for S (T)-polarization. We also calibrate the duration of the adiabatic spin transfer $w_R$. Fig. 3c shows the maximum nuclear field change $B_{max}$ reachable as a function of $w_R$, where both S- and T- polarizations are ineffective for short $w_R$ because of negligible adiabatic transfer probability $P_{LZ}$ [2,43]. $|B_{max}|$ reaches a maximum around $w_R \sim 0.8$ $\mu$s, after which the maximum efficiency is retained for the S-polarization sequence. In the case of T-polarization, however, for long $w_R$, $|B_{max}|$ decreases because of $D_T$ relaxation during the adiabatic passage.

By tuning $\delta R$ via the dc gate voltages and performing similar S-polarization experiments, we find that $B_{max}$ decreases with increasing $\delta R$ (Fig. 3d, see Extended Data Fig. 1). As is discussed subsequently, we find that the nuclear diffusion time scale exceeds 60 s regardless of $\delta R$, but the Overhauser field change per electron flip $b_0$ is strongly suppressed with increasing $\delta R$. Ultimately, the observation implies that the pulsed-gate-based nuclear control becomes inefficient in the non-interacting regime.

Returning to the condition $\delta R \sim 0.9$ $h \cdot$GHz, we demonstrate on-demand nuclear field programming. Fig. 3e (3f) shows the result of optimized S (T)-polarization with $t_{evol} = 0$ ns, $w_R = 1000$ ns ($t_{evol} = 0.62$ ns, $w_R = 600$ ns). Although the local fluctuations of the nuclear spins lead to random drift of the anti-crossing positions without the polarization pulse, $B_{nuc}$ builds toward (opposite to) the $B_0$ direction faster than the nuclear spin diffusion timescale when the polarization pulse is applied before each line scan. $\delta B_{nuc}$ rises to $B_{max}$ 80 mT (−40 mT) until a dynamic equilibrium is reached. Because only the $S_Z = 1/2$ states contribute to the T-polarization, $|B_{max}|$ for the T-polarization is about half of that for the S-polarization, implying that the state initialize to both $S_Z$ states with nearly equal probability at the EST position.

We also demonstrate bidirectional DNP by adjusting $t_{evol}$ in Fig. 3g. Fig. 3h illustrates the programming of $B_{nuc}$ by adjusting the adiabatic sweep amplitude $A_R$ of the S-polarization sequence. Because $B_{nuc}$ builds in the $B_0$ direction and drives the anti-crossing to deeper $\varepsilon$

(more to (1,2) charge configuration) under the S-polarization, $A_R$ serves as the limiting factor of $B_{max}$. Thus, a self-limiting DNP protocol can be realized.

Using a simple rate equation, we simulate the polarization-probe sequence (red-dashed curve in Fig. 3e, see Methods and Supplementary Note 3) and obtain $\tau_N \sim 62$ s and $b_0 \sim 2.58$ $h \cdot kHz \cdot (g^* \mu_B)^{-1}$ from the fit. In contrast, the DNP effect is negligible in our device with the two-electron $ST_0$ qubit[8] under the same repetition rate as in the WM regime (see Supplementary Note 4). Through optimization of the magnitude and direction of $B_0$, $b_0 \sim 3$ $h \cdot kHz \cdot (g^* \mu_B)^{-1}$ can be achieved with an $ST_0$ qubit in GaAs[2,8]. However, the obtained result shows that robust nuclear control can be achieved with WMs even in the regime where the same level of control cannot be achieved with an $ST_0$ qubit. In addition, residual polarization ~21.5 mT exists after turning off the polarization sequence (Fig. 3e), which diffuses within ~30 min. The large Knight shift gradient originating from the non-uniformly broadened WM wavefunction may be a possible cause of the long $\tau_N$. However, the newly observed phenomena in this study, including the dependence of $b_0$ on the tuning condition, require further investigations[44,45].

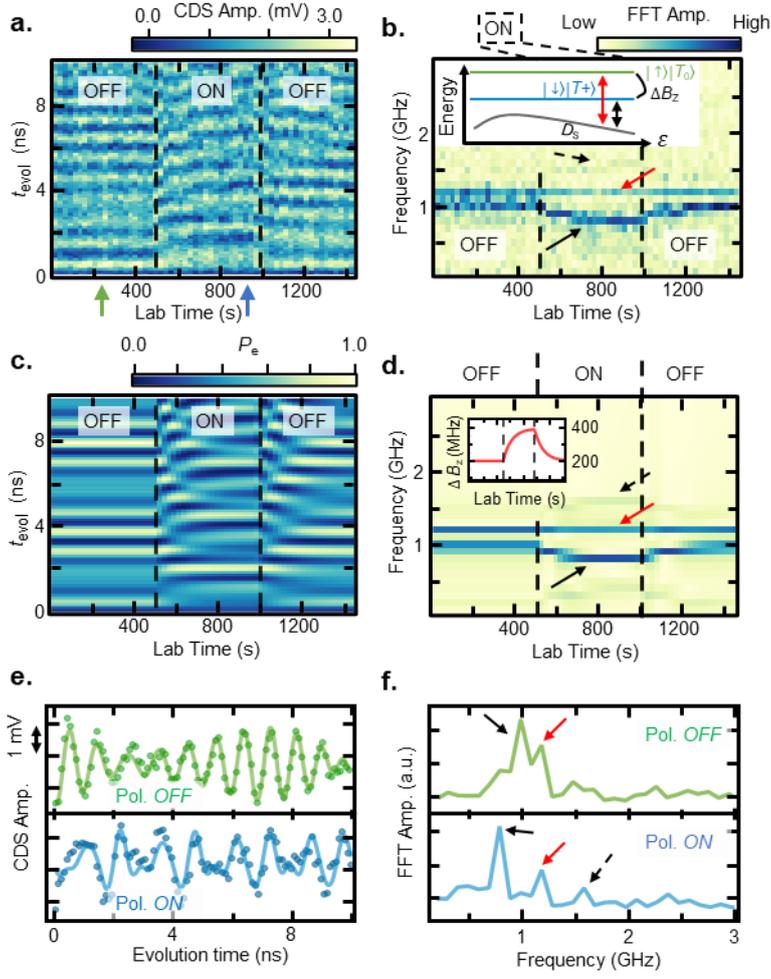

**Fig. 4.**

Furthermore, the WM's coherent LZS dynamics provide a novel approach to measure the spatial Overhauser field gradient $\Delta B_Z$ between QDs. When $\Delta B_Z$ is larger than the exchange splitting between $D_T(1,2;1/2)$ ($D_T(1,2;-1/2)$) and $Q(1,2;1/2)$ ($Q(1,2;-1/2)$), the eigenstates are expected to become $D_{T1}(1,2;1/2) = |\downarrow\rangle|T+\rangle$ ($D_{T1}(1,2;-1/2) = |\uparrow\rangle|T-\rangle$) and $D_{T0}(1,2;1/2) = |\uparrow\rangle|T_0\rangle$ ($D_{T0}(1,2;-1/2) = |\downarrow\rangle|T_0\rangle$)[46]. Because both states can tunnel-couple to $D_S(1,2;1/2)$ ($D_S(1,2;-1/2)$), the LZS oscillation reveals the $D_{T1} - D_S$ and $D_{T0} - D_S$ energy splittings. As can be inferred from the Hamiltonian (see Supplementary Note 5), although the $D_{T0} - D_S$ splitting is independent of the $\Delta B_Z$ and $B_Z$, the $D_{T1} - D_S$ splitting is modulated by $\Delta B_Z$ depending on the sign of $\Delta B_Z$ and $S_Z$, providing the direct measure of $\Delta B_Z$. Because the states can initialize to both $D_S(1,2;1/2)$ and $D_S(1,2;-1/2)$ at the EST position, the LZS oscillation captures the dynamics of both $S_Z = 1/2$ and $S_Z = -1/2$ subspaces.

Fig. 4a (4b) illustrates the LZS oscillation measurement of the WM multiplet states at $B_0 = 230$ mT in the time (frequency) domain with the S-polarization turned on and off at specific laboratory times. The FFT spectrum exhibits three different branches corresponding to the $D_{T0} - D_S$ (red arrow) and $D_{T1} - D_S$ (black and black-dashed arrows) where the beating patterns vary as the S-polarization induces changes in $\Delta B_z$. Two different $D_{T1} - D_S$ branches correspond to different $S_Z$ subspaces, where the sign of $\Delta B_z$ should be known to distinguish the $S_Z$ for each branch. The $D_{T0} - D_S$ splitting is the same for both $S_Z$ subspaces and is displayed as a single branch (red arrow). Fig. 4c (4d) shows the simulated time (frequency) domain signal of the same LZS oscillation, which agrees well with the experimental result (see Supplementary Note 6). As expected, the $D_{T0} - D_S$ splitting is constant regardless of $\Delta B_z$, whereas the $D_{T1} - D_S$ splitting is modulated along the polarization sequence.

The $D_{T0} - D_{T1}$ splitting without the polarization sequence implies the built-in $\Delta B_z \sim 200$ $h \cdot \text{MHz} \cdot (g^* \mu_B)^{-1}$ (35 mT), which is also confirmed by the ST$_0$ oscillation (see Supplementary Note 4). $\Delta B_z$ increases to 400 $h \cdot \text{MHz} \cdot (g^* \mu_B)^{-1}$ (70 mT) with the S-polarization and decreases to 200 $h \cdot \text{MHz} \cdot (g^* \mu_B)^{-1}$ after turning the polarization off. Thus, we conclude that the S-polarization yields the asymmetric pumping effect ($\Delta B_{\text{nuc}} \sim 200$ $h \cdot \text{MHz} \cdot (g^* \mu_B)^{-1}$) about the QD sites, whereas the $\Delta B_{\text{nuc}}$ direction can be experimentally checked, for example, via single-spin electric-dipole spin resonances[42]. Furthermore, the $D_{T0} - D_S$ splitting comprises the decoherence-free subspace for the qubit operations resilient to magnetic noises, where the coherent microwave control combined with the large polarization may enable leakage-free and state-selective transitions.

The present work uncovers the spin and energy structure of the WM states and explores the central-spin problem with strongly correlated WM states in semiconductor QDs. With the energy splitting of the WM ~ 0.9 $h$·GHz, we confirm the programmable DNP of $B_{nuc}$ ($\Delta B_{nuc}$) reaching (but not limited to) 80 mT (35 mT) via leakage spectroscopy and LZS oscillations. The $\tau_N$ exceeds 60 s, which, together with bidirectional polarizability, is beneficial for stabilizing the nuclear bath fluctuation and realizing long-lived nuclear polarization[10,15].

We anticipate several directions for further developments and applications of WM-enabled DNP. Similar experiments with a larger $\delta L/T_e$ ratio can enable high-fidelity single-shot readout for a faster measurement of the dynamics of nuclear polarization. This would further enable feedback loop control[10] and tracking[12,47] of nuclear environments in multielectron QDs. The real-time Hamiltonian estimation also improves frequency resolution for measuring instantaneous $\Delta B_{nuc}$, which may enable measurements of the degree of spatial localization within WMs. Furthermore, DNP becomes inefficient with increasing $E_{ST}$ of the WM, as discovered herein. This implies that the pulsed-gated electron-nuclear flip-flop probability is a strong function of the Wigner parameter, the microscopic origin of which requires more rigorous investigations.

**Methods**

**Device fabrication**

A quadruple QD device was fabricated on a GaAs/AlGaAs heterostructure with a 2DEG formed ~70 nm below the surface. The transport property of the 2DEG showed mobility $\mu = 2.6\times10^6$ cm$^2$(V·s)$^{-1}$ with electron density $n = 4.0\times10^{11}$ cm$^{-2}$ at temperature $T = 4$ K. Electronic mesa around the QD site was defined by the wet etching technique, and thermal diffusion of a metallic stack of Ni/Ge/Au was used to form the ohmic contacts. The depletion gates were deposited on the surface using standard e-beam lithography and metal evaporation of 5 nm Ti/30 nm Au. The lithographical width of the inner QD along the QD axis direction was designed to be ~10% wider than the outer dot to facilitate WM formation. The QD array was aligned to the [110] crystal axis, as shown in Fig. 1a. Although the magnetic field $B_0$ was intended to be applied perpendicular to the [110] axis to minimize the effect of spin-orbit interaction[2], the angular deviation was not strictly calibrated.

**Measurement**

The device was placed on a ~ 40 mK plate in a commercial dilution refrigerator (Oxford instruments, Triton-500). Ultra-stable dc-voltages were generated by battery-powered dc-sources (Stanford Research Systems, SIM928). They were then combined with rapid voltage pulses from an arbitrary waveform generator (AWG, Keysight M8195A with a sample rate up to 65 GSa/s) via homemade wideband ($10^1$–$10^{10}$ Hz) bias tees to be applied to the metallic gate electrodes. An LC-tank circuit with a resonant radio frequency (rf) of ~120 MHz was attached to the ohmic contact near the SET charge sensor to enable high-bandwidth ($f_{\rm BW}$ > 1 MHz) charge detection[27,30–33]. The reflected rf-signal was first amplified by 50 dB using two-stage low-noise cryo-amplifiers (Caltech Microwave Research, CITLF2 ×2 in series) at a 4 K plate. Next, it was further amplified by 25 dB at room temperature using a homemade low-noise rf-amplifier. The signal was then demodulated by an ultra-high-frequency lock-in amplifier (Zurich Instruments, UHFLI), which was routed to the boxcar integrator built in the

UHFLI. Trigger signals with a repetition period of 51 $\mu$s were generated by a field-programmable-gate array (FPGA, Digilent, Zedboard) to synchronize the timing of the AWG and the boxcar integrator for the CDS[33].

**Eigenstates of three-electron spin states**

Three-electron spin-multiplet structure consists of eight different eigenstates, which are four quadruplet states $Q(S_Z = 3/2)$, $Q(S_Z = 1/2)$, $Q(S_Z = -1/2)$, and $Q(S_Z = -3/2)$ and four doublet states $D_S(S_Z = 1/2)$, $D_T(S_Z = 1/2)$, $D_S(S_Z = -1/2)$, and $D_T(S_Z = -1/2)$, as shown in Table 1[46,48,49].

**Table 1**. **Three-electron spin states**

| State | Spin structure |
|---|---|
| $Q(S_Z = 3/2)$ | $\|\uparrow\uparrow\uparrow\rangle$ |
| $Q(S_Z = 1/2)$ | $\frac{1}{\sqrt{3}}(\|\uparrow\uparrow\downarrow\rangle + \|\uparrow\downarrow\uparrow\rangle + \|\downarrow\uparrow\uparrow\rangle)$ |
| $Q(S_Z = -1/2)$ | $\frac{1}{\sqrt{3}}(\|\downarrow\downarrow\uparrow\rangle + \|\downarrow\uparrow\downarrow\rangle + \|\uparrow\downarrow\downarrow\rangle)$ |
| $Q(S_Z = -3/2)$ | $\|\downarrow\downarrow\downarrow\rangle$ |
| $D_S(S_Z = 1/2)$ | $\frac{1}{\sqrt{2}}(\|\uparrow\uparrow\downarrow\rangle - \|\uparrow\downarrow\uparrow\rangle)$ |
| $D_T(S_Z = 1/2)$ | $\frac{1}{\sqrt{6}}(\|\uparrow\uparrow\downarrow\rangle + \|\uparrow\downarrow\uparrow\rangle - 2\|\downarrow\uparrow\uparrow\rangle)$ |
| $D_S(S_Z = -1/2)$ | $\frac{1}{\sqrt{2}}(\|\downarrow\downarrow\uparrow\rangle - \|\downarrow\uparrow\downarrow\rangle)$ |
| $D_T(S_Z = -1/2)$ | $\frac{1}{\sqrt{6}}(\|\downarrow\downarrow\uparrow\rangle + \|\downarrow\uparrow\downarrow\rangle - 2\|\uparrow\downarrow\downarrow\rangle)$ |

When $B_0 = 0$ T, the $D_S$ states, $D_T$ states, and $Q$ states are degenerate respectively, resulting in three different branches in the energy dispersion. We use a simple toy-model Hamiltonian adopted from the double QD hybrid qubit[35,36], which leads to a 6 × 6 Hamiltonian with the charge states considered as below. The ordered basis for the Hamiltonian is [$D_S(2,1)$, $D_T(2,1)$, $Q(2,1)$, $D_S(1,2)$, $D_T(1,2)$, $Q(1,2)$], where $n$ ($m$) denotes the number of electrons in the left (right) QD by ($n, m$).

$$H_{elec} = \begin{bmatrix} \varepsilon/2 & 0 & 0 & t_1 & -t_2 & 0 \\ 0 & \varepsilon/2 + \delta L & 0 & -t_3 & t_4 & 0 \\ 0 & 0 & \varepsilon/2 + \delta L & 0 & 0 & t_4 \\ t_1 & -t_3 & 0 & -\kappa \varepsilon/2 + \delta R & 0 & 0 \\ -t_2 & t_4 & 0 & 0 & -\varepsilon/2 + \delta R & 0 \\ 0 & 0 & t_4 & 0 & 0 & -\varepsilon/2 + \delta R \end{bmatrix} - (1)$$

Here, $\varepsilon$ is the energy detuning between the double QD, $t_i$ is the tunnel coupling strength between different orbitals (i = 1, 2, 3, 4), and $\delta L$ ($\delta R$) is the orbital energy splitting in the left (right) dot. Further, $\kappa$ is a factor to account for the different lever-arms of the ground and excited states in the (1,2) WM states[50], recently shown to be the consequence of many-body effects[28,29]. The Hamiltonian is utilized to obtain the energy spectra shown in Fig. 1. As we discuss in detail in Supplementary Note 6, the LZS oscillation at non-zero $B_0$ is simulated by adding the hyperfine interaction terms[46,48] to the aforementioned Hamiltonian and by solving the time-dependent Schrodinger equation with the experimentally obtained parameters.

**Rate equation**

Nuclear spin polarization and the diffusion process were phenomenologically modeled using a rate equation:

$$\frac{dB_{nuc}}{dt} = -B_{nuc}/\tau_N + b_0 P_{flip}/T_{rep}, \quad (2)$$

where $\tau_N$ is the nuclear spin diffusion time, $b_0$ is the Overhauser field change per electron spin-flip, $P_{flip}$ is the nuclear spin flop probability obtained from the Landau–Zener transition probability $P_{LZ}$ and the false initialization probability (see Supplementary Note 3), and $T_{rep}$ is the pulse repetition period. Using Eq. (2), we simulated the polarization–probe sequence shown in Fig. 3 with the experimental parameters including the time required for the amplitude sweep in the leakage probe step.

**Data availability**

The data that support the findings of this study are available from the corresponding author upon request.

**Acknowledgments**

This work was supported by the National Research Foundation of Korea (NRF) grant funded by the Korean Government (MSIT) (No. 2018R1A2A3075438, No. 2019M3E4A1080144, No. 2019M3E4A1080145, and No. 2019R1A5A1027055), Korea Basic Science Institute (National Research Facilities and Equipment Center) grant funded by the Ministry of Education (No. 2021R1A6C101B418), and Creative-Pioneering Researchers Program through Seoul National University (SNU). The cryogenic measurement used equipment supported by the Samsung Science and Technology Foundation under Project Number SSTF-BA1502-03. Correspondence and requests for materials should be addressed to DK ([dohunkim@snu.ac.kr](mailto:dohunkim@snu.ac.kr)).

**Author contributions**

DK and WJ conceived the project. WJ performed the measurements and analyzed the data. JK and HJ fabricated the device. JP, MC, HJ, and SS built the experimental setup and configured the measurement software. VU synthesized and provided the GaAs heterostructure. All the authors contributed to the preparation of the manuscript.

**Figure captions**

**Figure 1. Wigner molecule formation in a GaAs double quantum dot. a.** Scanning electron microscope image of a GaAs quantum dot (QD) device similar to the one used in the experiment. Green dots denote the double QD defined for Wigner molecule (WM) formation which is aligned along the [110] crystal axis (black arrow). The inner plunger gate $V_2$ is designed to have anisotropic confinement potential as shown in the right panel to facilitate the localization of the electronic ground state. Yellow circle: a radio-frequency (rf) single-electron transistor (rf-SET) charge sensor for rf-reflectometry. External magnetic field $B_0$ is applied along the direction denoted by the yellow arrow. **b.** Charge stability diagram of the double QD near the three-electron region spanned by $V_1$ and $V_2$ gate voltages. Green shaded region: the energy-selective tunneling (EST) position for the state readout and initialization. **c.** Landau–Zener–Stückelberg (LZS) oscillation of the WM at $B_0 = 0$ T. The relative phase evolution between the excited doublet ($D_T$) and the ground doublet ($D_S$) results in the oscillation captured by the EST readout. Red-dashed curve in the fast Fourier transformed (FFT) map shows energy dispersion calculated from the toy-model Hamiltonian. The calculation yields quenched orbital energy spacing of the inner dot $\delta R \sim 0.9\ h \cdot \text{GHz}$. **d.** Left (Right) panel: Energy spectrum along the (2,1)–(1,2) charge configuration in the non-interacting (strongly interacting, this work) regime with $\delta L \sim 100\ h \cdot \text{GHz}$ ($\delta L \sim 19\ h \cdot \text{GHz}$), and $\delta R \sim 100\ h \cdot \text{GHz}$ ($\delta R \sim 0.9\ h \cdot \text{GHz}$).

**Figure 2. Leakage spectroscopy and probabilistic nuclear polarization with the Wigner molecule. a.** Left panel: schematics of the energy levels for different external magnetic fields $B_0 > 0$ T. Crossings between different $S_Z$ states become anti-crossings aided by the transverse nuclear Overhauser field. Right panel: schematic of the pulse sequence for leakage spectroscopy and probabilistic dynamic nuclear polarization (DNP). The pulse diabatically drives the initialized $D_S(2,1;1/2)$ ($D_S(2,1;-1/2)$) to (1,2), and hold $\varepsilon$ for 100 ns $\gg T_2^*$. Upon

the coincidence of the pulse detuning and the anti-crossing, the state probabilistically evolves to $Q(1,2;3/2)$ ($Q(1,2;1/2)$) and flips the electron spin $\Delta m_S = +1$ which accompanies $\Delta m_N = -1$. **b.** Leakage spectroscopy of the Wigner molecule (WM) state as a function of $B_0$ and the pulse amplitude $A_P$. Black (Red) dotted curve shows the calculated energy splitting between $D_T$ ($Q$) and $D_S$ at $B_0 = 0$ T. Measurement-induced nuclear field shifts the dispersion opposite to the direction of $B_0$. **c. (d.)** Leakage measurement with an additional probabilistic polarization pulse with amplitude $A_P$' applied before each line sweep. The $A_P$' is fixed to 370 (450) mV, and the additional distortion in the leakage spectrum is shown as red circles near a pulse amplitude of 370 (450) mV. Black arrows denote the magnetic field sweep direction.

**Figure 3. Bidirectional and programmable dynamic nuclear polarization enabled by Wigner molecularization. a.** Top panel: Schematic of the anticrossings used for deterministic dynamic nuclear polarization (DNP). Bottom panel: pulse sequence used for S- and T-polarizations. For $t_{evol} = 0$ ns, the sequence corresponds to maximum S-polarization, which brings $D_S(1,2;1/2)$ ($D_S(1,2;-1/2)$) adiabatically across the anti-crossing to $Q(1,2;3/2)$ ($Q(1,2;1/2)$) flipping the electron spin with $\Delta m_S = +1$ and leading to $\Delta m_N = -1$ (blue arrow, S-polarization). For $t_{evol} = 600$ ns, the sequence corresponds to maximum T-polarization. Herein, the $D_T(1,2;1/2)$ prepared with a (Landau–Zener–Stückelberg) LZS-oscillation-induced $\pi$-pulse is adiabatically transferred to $D_S(1,2;1/2)$, resulting in $\Delta m_S = -1$ and $\Delta m_N = +1$ (red arrow, T-polarization), which has the opposite polarization effect compared to S-polarization. **b.** Change in the nuclear field $\delta B_{nuc}$ as a function of $t_{evol}$. The gray curve shows the corresponding LZS oscillation measurement reflecting the $D_T$ population. The $\delta B_{nuc}$ oscillates out of phase to the LZS oscillation owing to the oscillation of the S- and T-polarization ratio. **c.** The magnitude of the maximum polarization $B_{max}$ as a function of ramp time $w_R$. The $B_{nuc}$ saturates to $B_{max}$ when the polarization and the nuclear spin diffusion rate reach an equilibrium. For small $w_R$, the

$|B_{max}|$ decreases because of the small Landau–Zener transition probability $P_{LZ}$ for both S- (blue circle) and T-polarizations (red circle). In the case of T-polarization, $|B_{max}|$ decreases again for long $w_R$ owing to the lattice relaxation of the excited population. **d.** $B_{max}$ as a function of $\delta R$. The polarization gets more efficient for smaller $\delta R$ indicating a strong dependence of the nuclear polarization efficiency on the Wigner parameter. **e. (f.)** Dynamic nuclear control with the S (T)-polarization sequence. The red dotted line is the numerical fit derived from the simple rate equation-based model. The fit yields the nuclear spin diffusion time $\tau_N \sim 62$ s, with a polarization magnitude per spin flip of ~2.58 $h \cdot kHz \cdot (g^* \mu_B)^{-1}$. **g.** On-demand nuclear field programming via $t_{evol}$. **h.** Adiabatic ramp amplitude $A_R$ with $t_{evol} = 0$ ns realizing self-limiting nuclear field programming.

**Figure 4. Field gradient control and measurement.** Landau–Zener–Stückelberg (LZS) oscillation of the Wigner molecule (WM) states at $B_0 = 230$ mT in **a.** the time domain and **b.** the frequency domain with the S-polarization sequence. The oscillation reveals the relative phase oscillation of the $D_{T1} - D_S$ (black arrow, black dotted arrow) and $D_{T0} - D_S$ (red arrow) of both the $S_Z = 1/2$ and $S_Z = -1/2$ states. The $D_{T0} - D_S$ splitting is constant regardless of the magnetic field gradient $\Delta B_Z$, whereas the $D_{T1} - D_S$ energy spacing is modulated by the $\Delta B_Z$ depending on the sign of $\Delta B_Z$ and $S_Z$. The resultant beating is visible in **e. (f.)** the time (frequency) domain line-cut when the polarization is on (green arrow in **a.**) and off (blue arrow in **a.**). The line cuts in the time domain are numerically fitted to the sum of three sine functions (solid lines in **e.**) with different amplitudes. Three separate peaks are visible in the frequency domain **(f.)** when the $\Delta B_Z$ is largely polarized in the bottom panel (blue line) in **f**. Simulated LZS oscillation in **c.** the time domain and **d.** the frequency domain with the $\Delta B_Z$ in the inset of **(d.)**. The simulation in the frequency domain reproduces the branches shown in **(b.)**.

**Extended Data**

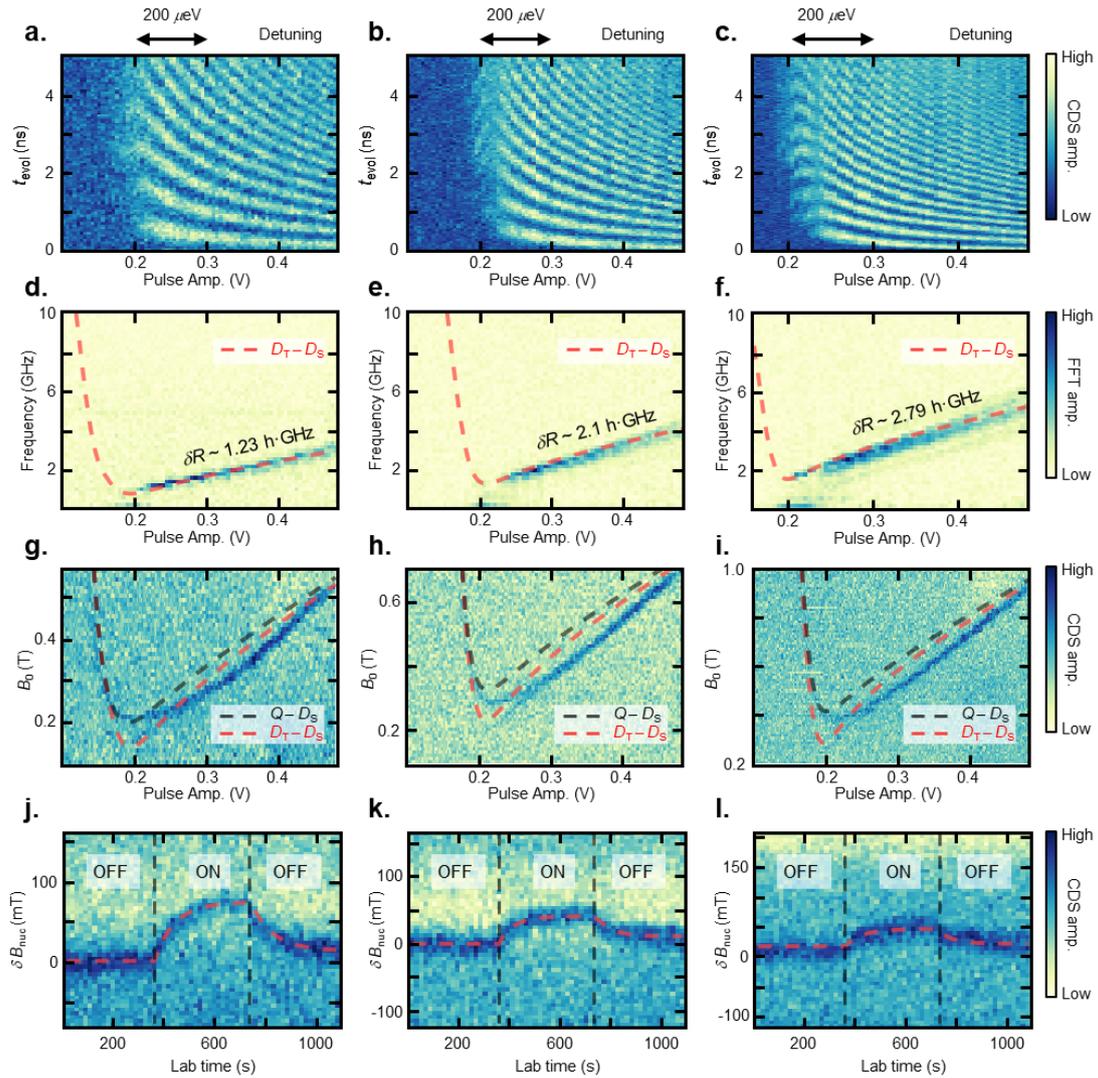

**Extended Data Fig. 1. Dynamic nuclear polarization under different tunings of Wigner molecule energy spectrum.** Time (frequency) domain Landau–Zener–Stuckelberg oscillation with the singlet-triplet splitting of the Wigner molecule (WM), $\delta R$ of **a.** (**d.**) ~ 1.23 $h \cdot$GHz, **b.** (**e.**) 2.1 $h \cdot$GHz, and **c.** (**f.**) 2.79 $h \cdot$GHz. Red-dashed curves in the frequency domain signals (**d.**, **e.**, and **f.**) show the energy splitting between $D_T$ and $D_S$ states derived from the toy-model Hamiltonian (see Methods section), from which we extract the magnitude of $\delta R$. The $\delta R$ is tuned with the dc-gate-voltages. Leakage spectroscopy of the WM with $\delta R$ of **g.** ~ 1.23 $h \cdot$GHz, **h.** 2.1 $h \cdot$GHz, and **i.** 2.79 $h \cdot$GHz. Red (black) dashed curves are the $D_T - D_S$ ($Q - D_S$) energy

spacings calculated from the toy-model Hamiltonian with the Lande g-factor $g^* \sim -0.4$. $\delta B_{nuc}$ measurement with the S-polarization turned on and off with $\delta R$ of **j.** $\sim 1.23\ h \cdot$ GHz, **k.** $2.1\ h \cdot$ GHz, and **l.** $2.79\ h \cdot$ GHz. Although the nuclear spin diffusion time is $\tau_N \sim 60$ s for all tuning, the nuclear polarization strength per electron spin flip $b_0$ decreases with increasing $\delta R$, as shown in Fig. 3d in the main text, resulting in smaller $B_{max}$ for larger $\delta R$ (i.e., smaller Wigner parameter)

**Supplementary Information**

**Supplementary Note 1. Electron temperature**

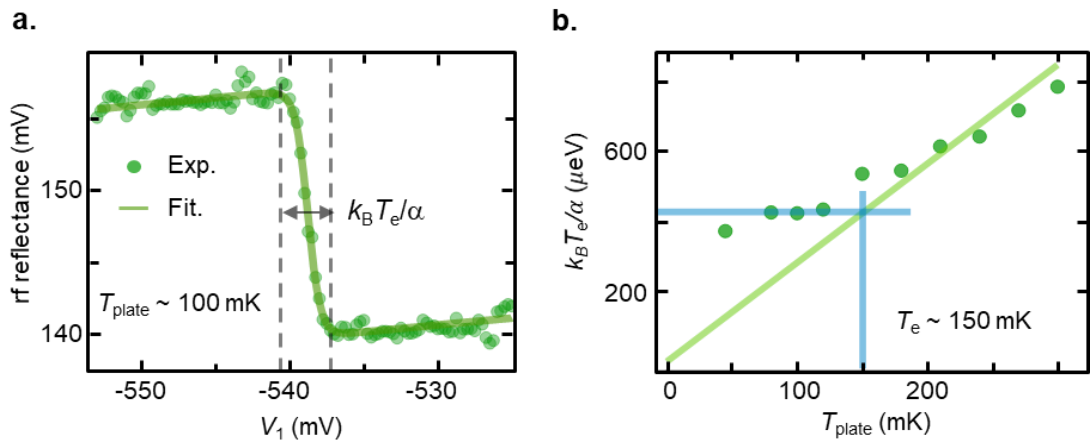

**Supplementary Figure S1. Electron temperature measurement. a.** Charge transition line broadening due to the finite electron temperature. The radio-frequency (rf)-single-electron transistor (rf-SET) charge sensing signal is recorded as a function of the gate voltage $V_1$ near the (2,1)–(1,1) charge transition at the mixing chamber plate temperature of the dilution refrigerator $T_{plate} \sim 100$ mK. The solid curve is a fit to the Fermi–Dirac distribution with a linear background slope, from which we obtain the thermal broadening $k_B T_e/\alpha$, where $k_B$ is the Boltzmann constant, $T_e$ is the electron temperature, and $\alpha$ is the lever arm of $V_1$. **b.** $k_B T_e/\alpha$ measured with varying $T_{plate}$. From the linear relationship for $T_{plate} > 200$ mK and plateau for $T_{mixing} < 100$ mK, we estimate $T_e = 150$ mK and $\alpha = 0.03$, respectively.

**Supplementary Note 2. Correlated double sampling (CDS)**

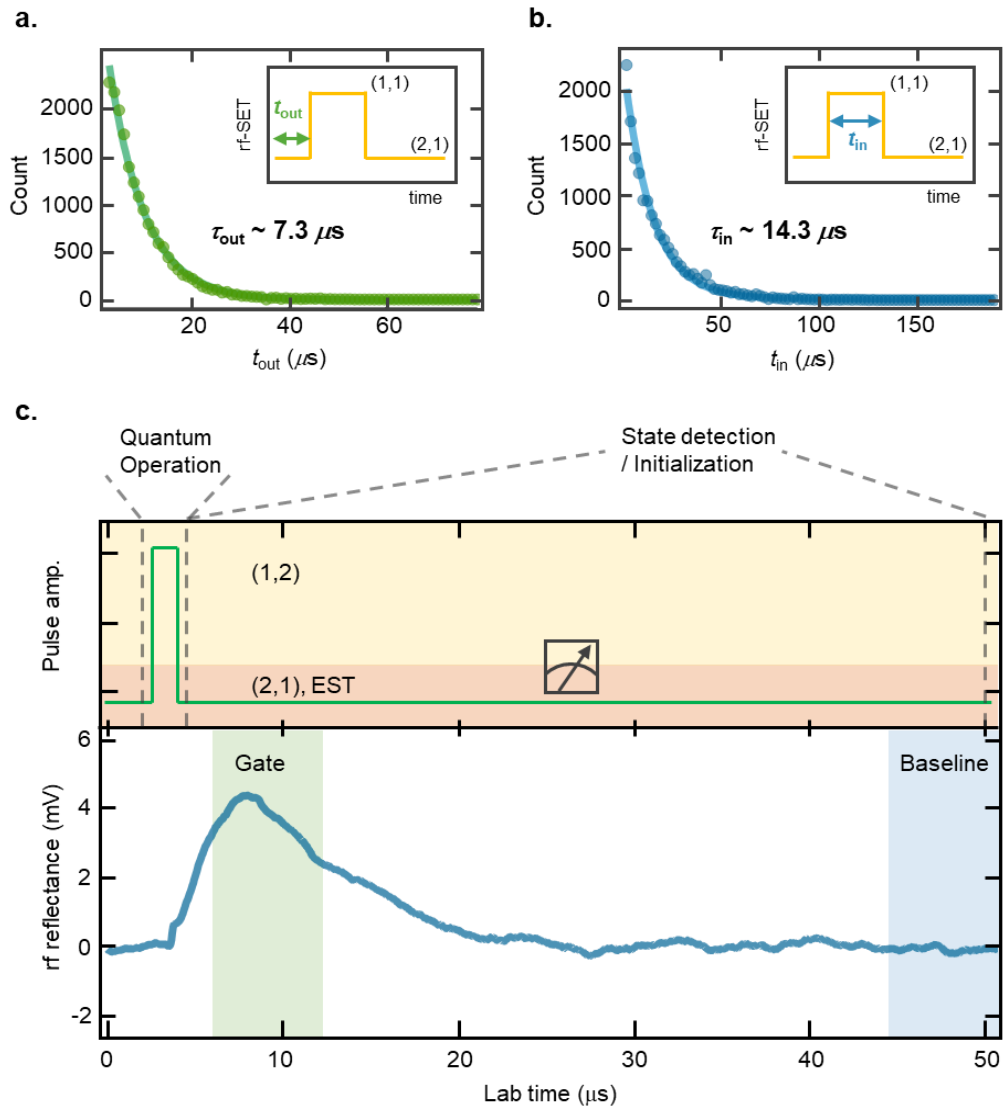

**Supplementary Figure S2. Tunneling time scale and correlated double sampling (CDS).**
**a. (b.)** Tunneling-out (-in) statistics. The solid curve is a fit to an exponential decay yielding the tunneling-out (-in) time $\tau_{out}$ ($\tau_{in}$) ~ 7 (14) $\mu$s. Inset in each figure shows a schematic of the charge sensor signal showing tunneling events in the (2,1) energy-selective tunneling (EST) region[1–4] recorded with the radio-frequency (rf)-single-electron transistor (rf-SET). **c.** Top panel: schematic of the quantum control sequence. The pulse brings the initialized state from (2,1) to the operation point in (1,2) and drives back to (2,1) for the EST readout and state initialization. Bottom panel: periodically averaged (~$10^6$ lines) ac-coupled rf-SET signal synchronized with the Landau–Zener–Stückelberg (LZS)-induced $X_\pi$ pulse. The dc-offset-eliminated CDS amplitude is generated by subtracting the baseline signal (blue shaded box) from the gate signal (green shaded box) and averaging ~$10^3$ times via the boxcar integrator. As discussed in the main text, the boxcar integration and the control waveform generation is synchronized to a trigger signal with period of 51 $\mu$s.

**Supplementary Note 3. Numerical simulation of the nuclear polarization sequence**

The nuclear field $B_\text{nuc}$ during the dynamic nuclear polarization (DNP) is numerically reproduced using the rate equation as follows:

$$\frac{dB_\text{nuc}}{dt} = -\frac{B_\text{nuc}}{\tau_N} + \frac{b_0 P_\text{flip}}{T_\text{rep}}. \quad (SE1)$$

As discussed in the Methods section, $\tau_N$ is the nuclear spin diffusion time, $T_\text{rep}$ is the repetition period of the polarization pulse, $b_0$ is the change in the nuclear field per electron spin flip, and $P_\text{flip}$ is the spin-flip probability obtained from the Landau–Zener transition probability $P_\text{LZ}$ and the false initialization probabilities $\beta$.

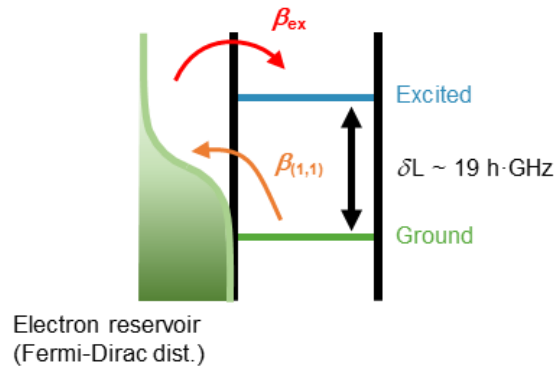

**Supplementary Figure S3. Schematic of false initialization at energy-selective tunneling.**
False initialization to the (1,1) ($\beta_{(1,1)}$, orange arrow) or the excited orbitals ($\beta_\text{ex}$, red arrow) may occur owing to thermal tunneling.

We first analyze the spin-flip probability $P_\text{flip}$ per adiabatic passage. Because of the small singlet-triplet splitting in the (2,1) EST region $\delta L \sim 19\ h\cdot\text{GHz}$, where $h$ is Planck's constant, the false initialization probability to (1,1) at the start of the pulse is $\beta_{(1,1)} \sim 0.37$ (Fig. S3, orange arrow), which does not contribute to the polarization. We also estimate the probability of the false initialization to the excited orbitals $\beta_{ex}$ from the Fermi–Dirac distribution with $T_e \sim 150$ mK, as described in Supplementary Note 1. With the Fermi level of the reservoir straddling in the middle of the $D_T - D_S$ splitting, we find $(\beta_{ex})^{-1} \sim Z = 1 + \exp(\delta L/2)/k_B T_e) = (0.049)^{-1}$, where $Z$ is the partition function[5]. Because the falsely initialized state in the excited orbital contributes to the polarization in the opposite direction, we calculate $P_{flip} = P_{LZ} \cdot (1 - \beta_{(1,1)} - 2\beta_{ex})$. We estimate $P_{LZ} \sim 0.5$ for the given adiabatic ramp width $w_R$ from Fig. 3, as the maximum efficiency is saturated for $w_R > 0.8$ $\mu$s. The resultant $P_{flip}$ is 0.26.

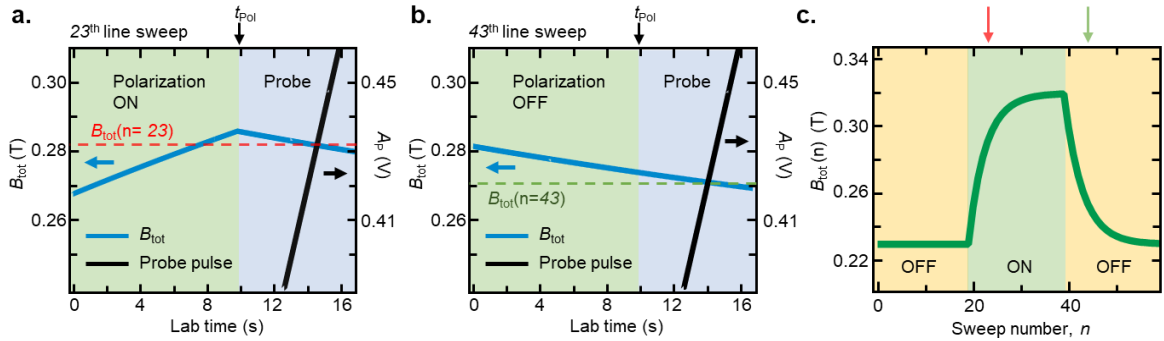

**Supplementary Figure S4. a.** Schematic of the net magnetic field $B_{tot} = B_0 + B_{nuc}$ during the polarization and the probe stages. During the polarization stage (green shaded area), $B_{nuc}$ builds up and then decays at the probe stage (blue shaded area) owing to nuclear diffusion. The decaying $B_{tot}$ is probed by the pulse amplitude sweep denoted by the black solid line. The crossing of $B_{tot}$ and the pulse amplitude is recorded as the leakage point (red-dotted line in **a.**, red arrow in **c.**). **b.** $B_{tot}$ with the polarization turned off. The crossing point (green-dotted line) is recorded as $B_{tot}$ ($n = 43$) shown in **c.** (green arrow). **c.** Simulation of $B_{tot}$ during the S-polarization sequence. When the polarization is turned on, $B_{tot}$ builds in the direction of the $B_0$ and then decays back when the polarization is turned off.

To simulate the polarization sequence, we consider the duration of the polarization stage $t_{Pol} \sim 10$ s and the adiabatic ramp amplitude $A_R$ by setting the $P_{flip}$ to 0.26 only if the laboratory time $t_{lab} < t_{Pol}$ and $B_{tot} = B_0 + B_{nuc} < B_L(A_R)$; otherwise, we set $P_{flip} = 0$. Here, $B_L$ is the one-to-one function between the pulse amplitude and the magnetic field strength obtained from the leakage spectrum in Fig. 2b. This reproduces the experimental situation where the polarization pulse is turned on for $t_{Pol}$ only when the anti-crossing is reachable with the maximum pulse amplitude, as shown in the green-shaded area in Fig. S4a. We convert the number of polarized nuclei to $B_{nuc}$ via $b_0$.

Based on the above-mentioned setting, we numerically mimic the leakage measurement shown in Fig. 3e. We check for the point where the crossing of the decaying $B_{tot}$ and the probe pulse amplitude (black line in Fig. S4a, S4b) occurs at the probe stage (Fig. S4a, (S4b), red (green) dashed line) and denote it as $B_{tot}(n)$ for the $n^{th}$ leakage measurement line sweep. Fig. S4c shows a collection of crossing points $B_{tot}(n)$ with the polarization turned on and off along $n$, which reflects the leakage measurement with the polarization sequence turned on and off, respectively. We fit $B_{tot}(n)$ to the leakage measurement in Fig. 3e and obtain $b_0 \sim$ 2.58 $h \cdot kHz \cdot (g^* \mu_B)^{-1}$ and $\tau_N \sim 62$ s.

# Supplementary Note 4. Inefficient nuclear polarization in the two-electron singlet-triplet qubit regime

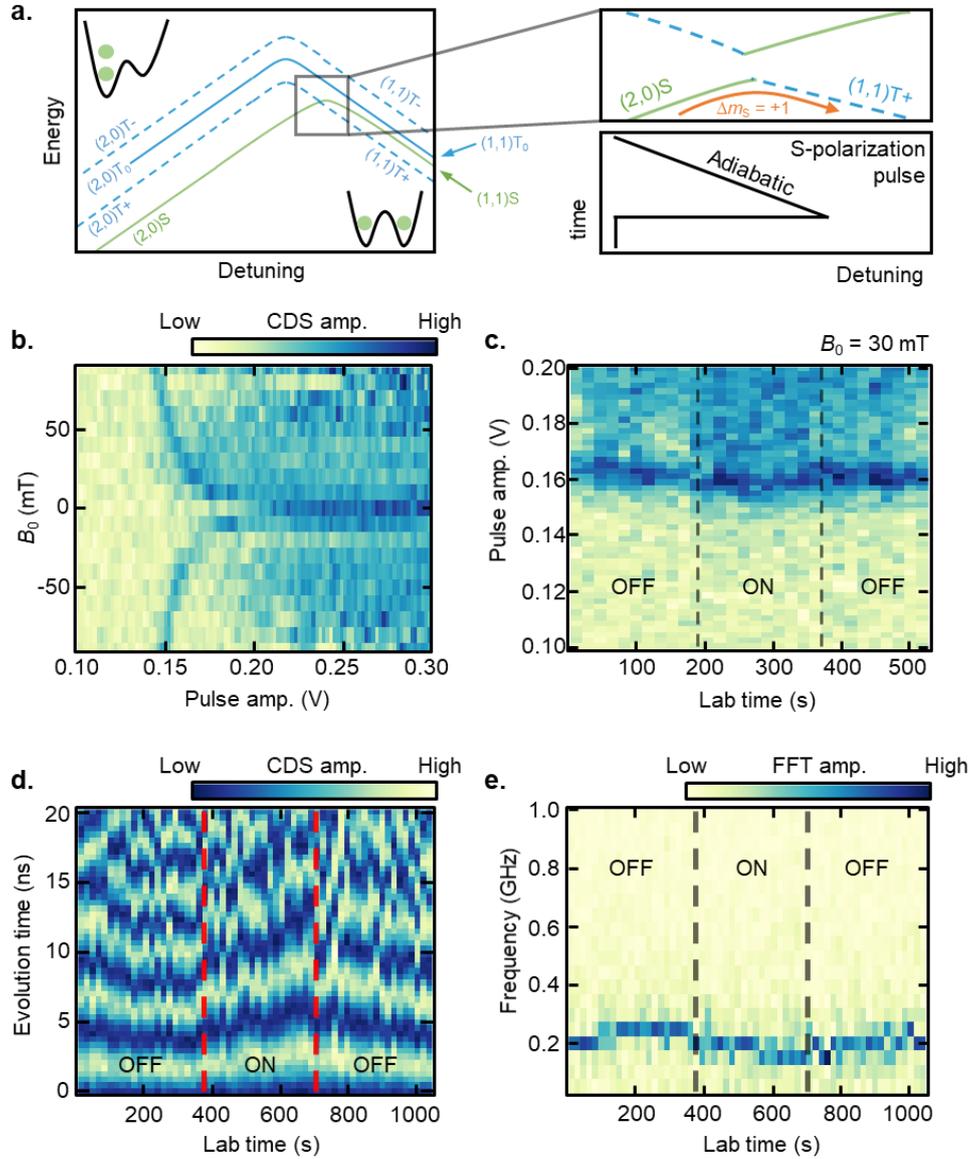

**Supplementary Figure S5. a.** Left panel: schematic of the singlet-triplet (ST$_0$) qubit energy levels in the two-electron regime. Zeeman-split T+ level crosses with the singlet branch (black rectangle) resulting in the Overhauser field-mediated anti-crossing. Right panel: magnified view of the anti-crossing with the pulse sequence shown below for the S-polarization ($\Delta m_S$ = +1, $\Delta m_N = -1$) with the ST$_0$ qubit[6,7]. **b.** Leakage spectroscopy (spin-funnel) of the singlet-triplet (ST$_0$) qubit. The spectrum reveals the S–T+ anti-crossing points as a function of $B_0$. **c.** Leakage measurement at $B_0$ = 30 mT with the S-polarization turned on and off with a pulse repetition period of 51 $\mu$s. No significant signature of $B_{nuc}$ exceeding the fluctuation was found. **d. (e.)** Time (frequency) domain signal of the ST$_0$ qubit Larmor oscillation at $B_0$ = 230 mT with the

S-polarization turned on and off. A built-in $|\Delta B_Z| = |B_Z^L - B_Z^R| \sim 200$ $h \cdot$MHz$\cdot(g^*\mu_B)^{-1}$ exists, where the additional polarization effect is not significantly larger than the fluctuation.

In this section, we show the two-electron singlet-triplet (ST$_0$) spin qubit operation to compare the nuclear polarization effect in the same device. Fig. S5a shows typical two-electron energy levels in a double quantum dot (QD)[8]. The Zeeman-split T+ level crosses with the singlet branch, and the crossing becomes an anti-crossing aided by the finite transverse nuclear Overhauser field[6,7] (right panel in Fig. S5a).

Utilizing the EST readout in the (2,0) charge configuration[3], we first measure the leakage spectrum of the ST$_0$ qubit by probing the S–T+ anti-crossings as a function of $B_0$ (Fig. S5b). Because the leakage position is sensitive to the magnetic field only for $|B_0| < 50$ mT, we set $B_0 = 30$ mT and investigate the effect of S-polarization in Fig. S5c. We use the same $T_{\text{rep}} \sim 51$ $\mu$s as described in the main text and measure the anti-crossing position with the polarization pulse turned on and off with the same polarization-probe sequence shown in Fig. 3e. As a result, we find that the polarization effect is found to not be as significant as in the Wigner molecule (WM) case shown in Fig. 3. This is consistent with a previous report[6], where a sizable $B_{\text{nuc}}$ is only observable for $T_{\text{rep}} < 30$ $\mu$s using ST$_0$ qubit.

The Larmor oscillation frequency of the ST$_0$ qubit corresponds to the size of the spatial magnetic field gradient $\Delta B_Z$ between the double QD (DQD)[8,9]. We also measure the ST$_0$ Larmor oscillation with the S-polarization turned on and off at $B_0 = 230$ mT, as shown in Fig. S5d, with the same sequence as in Fig. 4. We first note that there exists a built-in $\Delta B_Z \sim 200$ $h \cdot$MHz$\cdot(g^*\mu_B)^{-1}$ stemming from the nuclear Overhauser field, consistent with the energy splitting between the $D_{T1}$ and $D_{T0}$ energy levels without the polarization, as shown in Fig. 4b and 4d. In contrast to the WM case shown in the main text where the S-polarization yields a change of $|\Delta B_Z|=|B_L^Z - B_R^Z| \sim 200\ h\cdot\text{MHz}\cdot(g^*\mu_B)^{-1}$, the S-polarization with the ST$_0$ qubit does not induce a polarization that is significantly larger than the nuclear field fluctuation with the same $T_{\text{rep}} \sim 51\ \mu s$ [6]. This indicates that a large Knight field shift aided by the non-uniform broadening of the WM wavefunction may suppress the nuclear spin diffusion and lead to sizable nuclear polarization despite the slow pulse repetition rate[10,11].

**Supplementary Note 5. Magnetic Hamiltonian**

We adopt the hyperfine Hamiltonian from the exchange-only qubit defined in a triple QD[12] as follows. The ordered basis for the Hamiltonian is [$D_S(1/2)$, $D_T(1/2)$, $Q(1/2)$, $Q(3/2)$, $D_S(-1/2)$, $D_T(-1/2)$, $Q(-1/2)$, $Q(-3/2)$].

$$H_{\text{hf}} = \begin{bmatrix}
\frac{1}{2}B_{100}^Z & -\frac{1}{2\sqrt{3}}B_{01\bar{1}}^Z & -\frac{1}{\sqrt{6}}B_{01\bar{1}}^Z & \frac{1}{2\sqrt{2}}B_{01\bar{1}}^+ & -\frac{1}{2}B_{100}^- & \frac{1}{2\sqrt{3}}B_{01\bar{1}}^+ & -\frac{1}{2\sqrt{6}}B_{01\bar{1}}^- & 0 \\
-\frac{1}{2\sqrt{3}}B_{01\bar{1}}^Z & \frac{1}{6}B_{\bar{1}22}^Z & -\frac{1}{3\sqrt{2}}B_{\bar{2}11}^Z & \frac{1}{2\sqrt{6}}B_{\bar{2}11}^+ & \frac{1}{2\sqrt{3}}B_{01\bar{1}}^- & \frac{1}{6}B_{\bar{1}22}^- & -\frac{1}{2\sqrt{6}}B_{\bar{2}11}^- & 0 \\
-\frac{1}{\sqrt{6}}B_{01\bar{1}}^Z & -\frac{1}{3\sqrt{2}}B_{\bar{2}11}^Z & \frac{1}{6}B_{111}^Z & \frac{1}{2\sqrt{3}}B_{111}^+ & -\frac{1}{2\sqrt{6}}B_{01\bar{1}}^+ & -\frac{1}{2\sqrt{6}}B_{\bar{2}11}^+ & \frac{1}{3}B_{111}^- & 0 \\
\frac{1}{2\sqrt{2}}B_{01\bar{1}}^- & \frac{1}{2\sqrt{6}}B_{\bar{2}11}^- & \frac{1}{2\sqrt{3}}B_{111}^- & \frac{1}{2}B_{111}^Z & 0 & 0 & 0 & 0 \\
-\frac{1}{2}B_{100}^+ & \frac{1}{2\sqrt{3}}B_{01\bar{1}}^+ & -\frac{1}{2\sqrt{6}}B_{01\bar{1}}^- & 0 & -\frac{1}{2}B_{100}^Z & \frac{1}{2\sqrt{3}}B_{01\bar{1}}^Z & \frac{1}{\sqrt{6}}B_{01\bar{1}}^Z & \frac{1}{2\sqrt{2}}B_{01\bar{1}}^- \\
\frac{1}{2\sqrt{3}}B_{01\bar{1}}^- & \frac{1}{6}B_{\bar{1}22}^+ & -\frac{1}{2\sqrt{6}}B_{\bar{2}11}^- & 0 & \frac{1}{2\sqrt{3}}B_{01\bar{1}}^Z & -\frac{1}{6}B_{\bar{1}22}^Z & \frac{1}{3\sqrt{2}}B_{\bar{2}11}^Z & \frac{1}{2\sqrt{6}}B_{\bar{2}11}^- \\
-\frac{1}{2\sqrt{6}}B_{01\bar{1}}^+ & -\frac{1}{2\sqrt{6}}B_{\bar{2}11}^+ & \frac{1}{3}B_{111}^+ & 0 & \frac{1}{\sqrt{6}}B_{01\bar{1}}^Z & \frac{1}{3\sqrt{2}}B_{\bar{2}11}^Z & -\frac{1}{6}B_{111}^Z & \frac{1}{2\sqrt{3}}B_{111}^- \\
0 & 0 & 0 & 0 & \frac{1}{2\sqrt{2}}B_{01\bar{1}}^+ & \frac{1}{2\sqrt{6}}B_{\bar{2}11}^+ & \frac{1}{2\sqrt{3}}B_{111}^+ & -\frac{1}{2}B_{111}^Z
\end{bmatrix} \quad (SE2)$$

Here, $B_{abc}^r = aB_1^r + bB_2^r + cB_3^r$, where $r = z, +, -$, $B_d$ denotes the magnetic field on the $d^{\text{th}}$ electron, and $\bar{n} = -n$. The transverse magnetic field $B^+$ and $B^-$ couple different $S_Z$ subspaces with $|\Delta m_S| = 1$, where $S_Z$ is the spin projection to the quantization axis. Note that the spin-flip terms corresponding to $|\Delta m_S| = 2$ are not present.

For the LZS oscillation simulation shown in Fig. 4c, 4d, we assume that 1) the transverse Overhauser field $B^+$ and $B^-$ are negligibly small compared to $B^Z$, and 2) the spatial magnetic field gradient within a single QD is insignificant compared to that between the left and right QDs. In the (1,2) charge configuration in a DQD, we use $B_L = B_{d=1}$ to denote the magnetic field on the electron in the left QD and $B_R = B_{d=2} = B_{d=3}$ to denote the magnetic field experienced by the two electrons inside the right QD. Based on the notation and the two assumptions above, $H_{hf}$ can be simplified as (SE3).

$$H_{hf} = \begin{bmatrix} \frac{1}{2}B_L^Z & 0 & 0 & 0 & 0 & 0 & 0 & 0 \\ 0 & \frac{1}{6}(4B_R^Z - B_L^Z) & \frac{2}{3\sqrt{2}}\Delta B_Z & 0 & 0 & 0 & 0 & 0 \\ 0 & \frac{2}{3\sqrt{2}}\Delta B_Z & \frac{1}{6}(B_L^Z + 2B_R^Z) & 0 & 0 & 0 & 0 & 0 \\ 0 & 0 & 0 & \frac{1}{2}(B_L^Z + 2B_R^Z) & 0 & 0 & 0 & 0 \\ 0 & 0 & 0 & 0 & -\frac{1}{2}B_L^Z & 0 & 0 & 0 \\ 0 & 0 & 0 & 0 & 0 & -\frac{1}{6}(4B_R^Z - B_L^Z) & -\frac{2}{3\sqrt{2}}\Delta B_Z & 0 \\ 0 & 0 & 0 & 0 & 0 & -\frac{2}{3\sqrt{2}}\Delta B_Z & -\frac{1}{6}(B_L^Z + 2B_R^Z) & 0 \\ 0 & 0 & 0 & 0 & 0 & 0 & 0 & -\frac{1}{2}(B_L^Z + 2B_R^Z) \end{bmatrix}$$

(SE3)

$$= \begin{bmatrix} H_{hf,P} & 0 \\ 0 & H_{hf,N} \end{bmatrix}$$

$H_{hf,P}$ ($H_{hf,N}$) is the Hamiltonian in the positive (negative) spin subspace, where $H_{hf,N} = -H_{hf,P}$ holds. Diagonalizing $H_{hf,P}$ results in Eq. (SE4) as shown below with the ordered basis [$D_S(1,2; 1/2)$, $D_{T0}(1,2; 1/2)$, $D_{T1}(1,2; 1/2)$, $Q(1,2; 3/2)$] [13]. Here, $D_{T0}(1,2; 1/2) = |\uparrow\rangle|T_0\rangle$ and $D_{T1}(1,2; 1/2) = |\downarrow\rangle|T_+\rangle$, as mentioned in the main text. Further, $n$ ($m$) denotes the electron number inside the left (right) QD by ($n$, $m$; $S_z$).

$$H_{\mathrm{hf,P}} = \begin{bmatrix} \frac{1}{2}B_{\mathrm{L}}^{Z} & 0 & 0 & 0 \\ 0 & \frac{1}{2}B_{\mathrm{L}}^{Z} & 0 & 0 \\ 0 & 0 & \frac{1}{2}B_{\mathrm{L}}^{Z} - \Delta B_{Z} & 0 \\ 0 & 0 & 0 & \frac{1}{2}(B_{\mathrm{L}}^{Z} + 2B_{\mathrm{R}}^{Z}) \end{bmatrix} \quad \text{(SE4)}$$

$D_{\mathrm{T1}} - D_{\mathrm{T0}}$ splitting is governed by $\Delta B_{Z}$, providing a direct measure of the size of the spatial magnetic field gradient, $\Delta B_{Z}$. We emphasize that $D_{\mathrm{T0}} - D_{\mathrm{S}}$ splitting is now independent of the magnetic field strength, providing a decoherence-free subspace for high-fidelity qubit operations. However, we note that $D_{\mathrm{T0}} - D_{\mathrm{S}}$ splitting may still be disturbed by the magnetic field gradient noise within the right QD, which we assume to be negligible compared to $\Delta B_{Z}$. After implementing the single-shot readout-based real-time Hamiltonian estimation technique[14], we anticipate that the investigation of the temporal dynamics of $D_{\mathrm{T0}} - D_{\mathrm{S}}$ splitting may enable the study of the magnetic field behavior within a single QD. This in turn would be helpful to reveal the spatial distribution of the WM wavefunction.

**Supplementary Note 6. Simulation of the Landau–Zener–Stückelberg oscillation**

As discussed in Supplementary Note 5, we neglect the transverse magnetic field contribution and do not consider the transition between different $S_{Z}$ subspaces in the LZS oscillation simulation. This allows us to analyze the dynamics of the $S_{Z} = 1/2$ and $S_{Z} = -1/2$ subspaces separately and ignore the $|S_{Z}| = 3/2$ subspace. We combine the reduced $S_{Z} = 1/2$ ($S_{Z} = -1/2$) hyperfine Hamiltonian with the electronic Hamiltonian shown in the Methods section to describe the dynamics in the $S_{Z} = 1/2$ ($S_{Z} = -1/2$) subspace. The hyperfine Hamiltonian in the $S_{Z} = 1/2$ subspace with the charge configurations is explicitly considered as follows. The ordered basis for the Hamiltonian is [$D_S(2,1; 1/2)$, $D_T(2,1; 1/2)$, $Q(2,1; 1/2)$, $D_S(1,2; 1/2)$, $D_T(1,2; 1/2)$, $Q(1,2; 1/2)$].

$$H_{hf,1/2} = \begin{bmatrix} \frac{1}{2}B_R^Z & 0 & 0 & 0 & 0 & 0 \\ 0 & \frac{1}{6}(4B_L^Z - B_R^Z) & -\frac{2}{3\sqrt{2}}\Delta B_Z & 0 & 0 & 0 \\ 0 & -\frac{2}{3\sqrt{2}}\Delta B_Z & \frac{1}{6}(B_R^Z + 2B_L^Z) & 0 & 0 & 0 \\ 0 & 0 & 0 & \frac{1}{2}B_L^Z & 0 & 0 \\ 0 & 0 & 0 & 0 & \frac{1}{6}(4B_R^Z - B_L^Z) & \frac{2}{3\sqrt{2}}\Delta B_Z \\ 0 & 0 & 0 & 0 & \frac{2}{3\sqrt{2}}\Delta B_Z & \frac{1}{6}(B_L^Z + 2B_R^Z) \end{bmatrix}$$ (SE 5)

For numerical reproduction of the LZS oscillation shown in Fig. 4c, we solve the time-dependent Schrödinger equation by varying the detuning parameter $\varepsilon$ according to the pulse shape. As the state probabilistically initializes to either $D_S(2,1; 1/2)$ or $D_S(2,1; -1/2)$ at EST, we simulate the LZS oscillations of the $S_Z = 1/2$ and the $S_Z = -1/2$ cases separately. The simulated oscillations are then averaged assuming the equal initialization probability to $D_S(2,1; 1/2)$ and $D_S(1,2; -1/2)$.